\documentstyle[12pt]{article}
\begin{document}

\begin{center}
\vspace*{1cm}{\Large {\bf
Left-handed materials in metallic magnetic granular composites }}
\medskip \\[0pt]
\smallskip S.T.Chui, Z. F. Lin and L.-B. Hu\\[0pt]
\smallskip

\vspace*{0.2cm} {\it Bartol Research Institute, University of Delaware,
Newark, Delaware, U.S.A} \vspace*{0.2cm}
\end{center}

\centerline{\bf\large Abstract} \vspace*{1cm}
There is recently interests in the ``left-handed '' materials.
In these materials
the direction of the wave vector of electromagnetic radiation
is opposite to the direction of the energy flow.
We present simple arguments that suggests that magnetic composites can also
be left-handed materials. However, the physics involved seems to be different
from the original argument. In our argument, the imaginary part of
the dielectric constant is much larger than the real part, opposite
to the original argument.

\medskip

 \vspace*{1cm}

There has been interest recently in left-handed
material \cite{Ves,Pendry,Smith}, in
which the direction of energy flow is opposite to the direction
of the wave vector. The original material consists of arrays of
rings and wires \cite{Pendry,Smith}.
We have recently proposed a different class of
material consisting of metallic magnetic granular
composites \cite{chuihu}.
Based on the effective medium approximation, we show
that by incorporating metallic magnetic nanoparticles into an appropriate
insulating matrix and controlling the directions of magnetization of
metallic magnetic components and their volume fraction, it may be possible
to prepare a composite medium of low eddy current loss which is
left-handed for electromagnetic waves propagating in some special direction
and polarization in a frequency region near the ferromagnetic resonance
frequency. In this paper, we present a simplified physical explanation
of our results.

We consider an
idealized metallic magnetic granular composite consisting of two types of
spherical particles,  in which  one type of particles are metallic
ferromagnetic grains of radius $R_{1}$,  the other type are non-magnetic
dielectric (insulating) grains of radius $R_{2}$.
For simplicity in this analysis we assume that $R_1=R_2$,
each grain is assumed to be homogeneous.
The volume fraction of the magnetic (nonmagnetic) particles is given by 
$f_1$ ($f_2$) with $f_1+f_2=1$.
The directions of magnetization of all metallic magnetic
grains are assumed to be in the same direction. For
length scales much larger than the grain sizes,  the composite can be
considered as a homogeneous magnetic system.  The permittivity and
permeability of non-magnetic dielectric grains are both scalars, and will be
denoted as $\epsilon _{1}$ and $\mu _{1}$. The permittivity of metallic
magnetic grains will be denoted as $\epsilon _{2}$ and will be taken to have
a Drude form: $\epsilon _{2}=1-\omega _{p}^{2}/\omega (\omega +i/\tau )$,
 where $\omega _{p}$ is the plasma frequency of the metal and $\tau $ is a
relaxation time.  Such a form of $\epsilon $ is representative of a variety
of metal composites \cite{bk82,bk78}.
The permeability of metallic magnetic grains
are second-rank tensors and will be denoted as $\hat{\mu}_{2}$,  which can
be derived from the Landau-Lifschitz equations \cite{7}.
Assuming that the
directions of magnetization of all magnetic grains are in the direction of \
the $z$-axis, $\hat{\mu}_{2}$ will have the following form \cite{7}:
\begin{equation}
\hat{\mu _{2}}=\left[
\begin{array}{cccc}
\mu _{a} & -i\mu ^{\prime } & 0 &  \\
i\mu ^{\prime } & \mu _{a} & 0 &  \\
0 & 0 & 1 &
\end{array}
\right]
\end{equation}
where \cite{err}
\begin{eqnarray}
\mu _{a} &=&1+\frac{\omega _{m}(\omega _{0}-i\alpha \omega )}{(\omega
_{0}-i\alpha \omega )^{2}-\omega ^{2}}, \\
\mu ^{\prime } &=&-\frac{\omega _{m}\omega }{(\omega _{0}-i\alpha \omega
)^{2}-\omega ^{2}}.
\end{eqnarray}
Here $\omega _{0}=\gamma \vec{H_{0}}$ is the ferromagnetic resonance
frequency, $H_{0}$ is the effective magnetic field in magnetic particles
and may be a sum of the external magnetic field,  the effective anisotropy
field and the demagnetization field;
$\omega _{m}=\gamma \vec{M_{0}}$,
with $\gamma $ the gyromagnetic ratio and $M_{0}$ the saturation
magnetization of magnetic particles;
$\alpha $ is the magnetic damping
coefficient; and, finally,
$\omega $\ is the frequency of incident electromagnetic waves.
Typical frequency range of the FMR resonance is controlled by the spin wave
energy at zero wave vector. This is of the order of GHz 
for such soft magnets as permalloy. 
The dispersion of the spin wave frequency is controlled 
by the exchange constant of the system.
We consider incident electromagnetic waves propagating in
the direction of the magnetization.

Our numerical calculation in Ref.\cite{chuihu} was performed without
assuming the long wavelength limit. The effective medium equation
is given by
\begin{eqnarray}
&&\sum_{i=1,2}f_{i}\sum_{l=1}^{\infty }(2l+1)[\frac{k_{eff}\psi _{l}^{\prime
}(k_{i}R_{i})\psi _{l}(k_{eff}R_{i})-k_{i}\psi _{l}(k_{i}R_{i})\psi
_{l}^{\prime }(k_{eff}R_{i})}{k_{eff}\psi _{l}^{\prime }(k_{i}R_{i})\zeta
_{l}(k_{eff}R_{i})-k_{i}\psi _{l}(k_{i}R_{i})\zeta _{l}^{\prime
}(k_{eff}R_{i})}  \nonumber \\
&&+\frac{k_{i}\psi _{l}^{\prime }(k_{i}R_{i})\psi
_{l}(k_{eff}R_{i})-k_{eff}\psi _{l}(k_{i}R_{i})\psi _{l}^{\prime
}(k_{eff}R_{i})}{k_{i}\psi _{l}^{\prime }(k_{i}R_{i})\zeta
_{l}(k_{eff}R_{i})-k_{eff}\psi _{l}(k_{i}R_{i})\zeta _{l}^{\prime
}(k_{eff}R_{i})}]=0,
\label{keff}
\end{eqnarray}
where $R_{i}$ is the radius of the $i$th type of grains,  and
\begin{eqnarray}
k_{1} &=&\omega \lbrack \epsilon _{1}\mu _{1}]^{1/2}, \\
k_{2} &=&\omega \lbrack \epsilon _{2}\mu _{2}^{(\pm )}]^{1/2}, \\
\psi _{l}(x) &=&xj_{l}(x), \\
\zeta _{l}(x) &=&xh_{l}^{(1)}(x),
\end{eqnarray}
with $j_{l}(x)$ and $h^{(1)}_{l}(x)$
the usual spherical Bessel and Hankel
functions. $\mu_{1}$ is the permeability of non-magnetic dielectric
grains, $\mu _{2}^{(+)}=\mu _{a} - \mu ^{\prime }$
$\mu_2^{(-)}=\mu_{a}+\mu^{\prime}$ are the effective permeability of
magnetic grains for positive and negative circularly polarized waves
respectively, and $f_1$ and $f_2$ are volume fractions of two types of
grains.

We first simplify these equations in the long wavelength limit, the
experimentally relevant situation.
In the long wavelength limit the $l=1$ terms in Eq.(\ref{keff}) dominate.
With the use of
$j_1\approx x/3$, $h_1\approx -i/x^2$,
$\psi_1\approx x^2/3$, $\zeta_1(x)\approx -i/x,$
$\psi_1'\approx x/1.5$, $\zeta_1(x)'\approx i/x^2$,
Eq.(\ref{keff}) can be simplified to:
\begin{equation}
\sum_{i=1,2}f_{i}\frac{ k_{eff}^2-k_{i}^2}{2k_{eff}^2 +k_{i}^2} =0.
\label{keffs}
\end{equation}
Since $k^2\propto \epsilon\mu$, we obtain the effective medium equation:
\begin{equation}
\sum_{i=1,2}f_{i}\frac{ \langle\mu\epsilon\rangle_{eff}-\mu_i\epsilon_i}
{2\langle\mu\epsilon\rangle_{eff} +\mu_i\epsilon_i} =0
\end{equation}
This is of the well known Bruggeman\cite{Bru} form that one is familair with.
We note that in general $\langle\mu\epsilon\rangle_{eff}\neq
\langle\mu\rangle_{eff}\langle\epsilon\rangle_{eff}$.

If the composite can truly be treated as a homogeneous magnetic system in
the case of grain sizes much smaller than the characteristic wavelength,
electric and magnetic fields in the composite should also be either positive
or negative circularly polarized and can be expressed as :
\begin{eqnarray}
\vec{E}(\vec{r},t) &=&\vec{E}_{0}^{(\pm )}e^{ikz-\beta z-i\omega t}
\label{Efld} \\
\vec{H}(\vec{r},t) &=&\vec{H}_{0}^{(\pm )}e^{ikz-\beta z-i\omega t}
\label{Hfld}
\end{eqnarray}
where $\vec{E}_{0}^{(\pm )}=E_0(\hat{x}\mp i\hat{y})$,
$\vec{H}_{0}^{(\pm )}=H_0(\hat{x}\mp i\hat{y})$,
$k={\rm Re}[k_{eff}]$ is the effective wave
number, $\beta ={\rm Im}[k_{eff}]$ is the effective damping coefficient
caused by the eddy current, and $k_{eff}=k+i\beta $
is the effective propagation constant.
In Eqs.(\ref{Efld}) and (\ref{Hfld})
the signs of $k$ and $\beta $ can both be
positive or negative depending on the
directions of the wave vector and the energy flow.
For convenience we assume that the direction of energy flow is in
the positive direction of the $z$ axis,
i.e., we assume $\beta >0$ in
Eqs.(\ref{Efld}) and (\ref{Hfld}),
but the sign of $k$ still can be positive or negative.
In this case,  if $k>0$, the phase velocity and energy
flow are in the same directions,  and
from Maxwell's equation,  one can
see that the electric and magnetic field $\vec{E}$ and $\vec{H}$ and the
wave vector
$\vec{k}$ will form a right-handed triplet of vectors.
This is the usual
case for right-handed materials. In contrast,  if $k<0$,
the phase velocity and energy flow are in opposite directions,
and $\vec{E}$, $\vec{H}$ and $\vec{k}$ will form a left-handed
triplet of vectors.  This is just
the peculiar case for left-handed materials.
So, for incident waves of a
given frequency $\omega$, we can determine whether wave propagations in the
composite is right-handed or left-handed through the relative sign changes
of $k$ and $\beta$.

For systems of experimental interest (such as composites of Ge, Si 
with Ni, permalloy, or Co), the modulus of the complex number
$\eta=(\mu_1\epsilon_1)/(\mu_2\epsilon_2)$ is much smaller than unity.
Eq.(\ref{keffs}) can then be solved explicitly
as a power of series in $\eta$ to give
\begin{equation}
k_{eff}^2=k_1^2 c_0^2(1-2c_1\eta)
\end{equation}
where
$$
c_0^2=\frac{1}{f_1-2f_2} \ \ \ \mbox{\rm and} \ \ \
c_1=\frac{9f_1f_2}{2(f_1-2f_2)^2}
$$
Note that $k_1^2$ is real for
the nonmagnetic dielectric grains while $\eta$ is complex
with its modulus $|\eta|<<1$.
To leading order in $\eta$, the effective wavevector $k_{eff}$ is real.
The imaginary part comes in in the next order of $\eta$.
Thus the loss of the effective medium is small when the fraction of 
the magnetic particles, $f_2<1/3$.
We next show that the sign of $Im(k_{eff})$ is controlled by $\mu_2$.

We write $\eta$ in its polar form as
\begin{equation}
\eta=|\eta|(\cos\phi-i\sin\phi)
\end{equation}
with
\begin{equation}
\cos\phi=\frac{\mu_{2r}\epsilon_{2r}-\mu_{2i}\epsilon_{2i}}{A}
    \approx\frac{-\mu_{2i}\epsilon_{2i}}{A}
 \ \ \ \mbox{\rm and} \ \ \
\sin\phi=\frac{\mu_{2r}\epsilon_{2i}+\mu_{2i}\epsilon_{2r}}{A}
    \approx \frac{\mu_{2r}\epsilon_{2i}}{A}
\end{equation}
where $A=|\mu_2\epsilon_2|$,
$\epsilon_{2r}={\rm Re}[\epsilon_2]$,
$\epsilon_{2i}={\rm Im}[\epsilon_2]$,
$\mu_{2r}={\rm Re}[\mu_2]$, $\mu_{2r}={\rm Im}[\mu_2]$,
and use has been made of
$\epsilon_{2i}>>\epsilon_{2r}$ and
$\mu_{2i}\sim \mu_{2r}$ for the metallic magnetic grains.
The effective wave number can be written as
\begin{equation}
k_{eff}= \pm k_1 c_0 \left(1
  + i \frac{\mu_{2r}\epsilon_{2i}}{A} c_1|\eta|\right)
\end{equation}
There are two solutions.
As we emphasized above, the direction of the energy flow is in the
positive direction of the $z$ axis, implying a positive imaginary part
of $k_{eff}$, i.e, we pick a solution so that with $k_{eff}=k+i\beta$,
$\beta>0$. Now $\mu_{2r}$ changes sign
when $\omega$  crosses $\omega_0$. So the solution that one picks
changes also.
More specifically, one has
\begin{equation}
k_{eff}=\left\{
\begin{array}{lllll}
 c_0 k_1 \left(1 + i \displaystyle\frac{\mu_{2r}\epsilon_{2i}}{A} c_1|\eta|
         \right) & &
(\omega < \omega_0, & \mu_{2r}>0) \\
-c_0 k_1 \left(1+i \displaystyle\frac{\mu_{2r}\epsilon_{2i}}{A} c_1|\eta|
         \right) & &
(\omega > \omega_0, & \mu_{2r}<0)
\end{array}
\right.
\end{equation}
The real part
of $k_{eff}$ should change sign to ensure the positive definite of
$\beta$.
This shows that in certain frequecy region $\omega>\omega_0$,
the wave vector is in the opppsotie direction of the energy flow.

In conclusion, we have here offered a simple physical explanation
for the left-handedness of magnetic composites. When the metallic concentration
is below the conduction percolation threshold, the imaginary part of the
effective response function is small. However, the sign
of this small imaginary term changes as the frequency crosses the ferromagnetic
resonance frequency. To insure energy conservation so that the sign
of the imaginary part of the effcetive wavevector remains the same, the
``branch'' of the solution changes. As a result, the real part
of the effective wave vector changes sign.
The original picture of the left-handed material focuses on the
{\bf real} part of the electric and magnetic susceptibilities.
Our explanation for our material focuses on the imaginary part of
the response functions. The physics in our case
seems to be different from the original picture.

This research is supported in part by the Army Research Lab
through the composite center at the University of Delaware.
We thank John Xiao for helpful discussion.


\begin{thebibliography}{10}
\bibitem{err} There are some typographical errors in signs
in the expression of $\mu_a$ and $\mu'$ in
Eqs.(2) and (3) of Ref.\cite{chuihu}.
\end{thebibliography}
\end{document}